\documentclass[aps,prl,floatfix,twocolumn]{revtex4-1}

\usepackage{amsmath,amssymb}
\usepackage[pdftex]{hyperref,graphicx}
\hypersetup{colorlinks = true, urlcolor = blue, linkcolor = blue, citecolor = blue}

\usepackage{physics}
\usepackage{xcolor}
\usepackage{bm}
\usepackage{nicefrac}

\begin{document}
	
	\title{Quantum phase diagram of a Moir\'e-Hubbard Model}

\author{Haining Pan}
\affiliation{Condensed Matter Theory Center and Joint Quantum Institute, Department of Physics, University of Maryland, College Park, Maryland 20742, USA}
\author{Fengcheng Wu}
\affiliation{Condensed Matter Theory Center and Joint Quantum Institute, Department of Physics, University of Maryland, College Park, Maryland 20742, USA}
\author{Sankar Das Sarma}
\affiliation{Condensed Matter Theory Center and Joint Quantum Institute, Department of Physics, University of Maryland, College Park, Maryland 20742, USA}

\begin{abstract}
	We theoretically study a generalized Hubbard model on moir\'e superlattices of twisted bilayers, and find very rich filling-factor-dependent quantum phase diagrams tuned by interaction strength and twist angle. Strong long-range Coulomb interaction in the moir\'e-Hubbard model induces Wigner crystals at a series of fractional filling factors. The effective lattice of the Wigner crystal is controlled by the filling factor, and can be triangle, rectangle, honeycomb, kagome, etc, providing a single platform to realize many different spin models on various lattices by simply tuning carrier density. In addition to Wigner crystals that are topologically trivial, interaction-induced Chern insulators emerge in the phase diagram. This finding paves a way for engineering interaction-induced quantum anomalous Hall effect in moir\'e-Hubbard systems where the corresponding single-particle moir\'e band is topologically trivial.
\end{abstract}

\maketitle

\textit{Introduction.} --- Twisted bilayers with a long-period moir\'e pattern provide versatile platforms to study strongly correlated physics, as many-body interactions are effectively enhanced in narrow moir\'e bands. It has been theoretically proposed that a generalized Hubbard model can be simulated in twisted bilayers based on  group-VI transition metal dichalcogenides (TMDs) \cite{wu2018hubbard, wu2019topological}, which have fewer low-energy degrees of freedom compared to twisted bilayer graphene \cite{bistritzer2011moirea, cao2018unconventional,cao2018correlated} and therefore, allow quantum simulations of model Hamiltonians.  Recent experiments \cite{regan2020mott,tang2020simulation,wang2020correlated,xu2020abundance,jin2020stripe,huang2020correlated} performed using a variety of techniques on twisted bilayer TMDs found compelling  evidence of correlated insulators (CIs) not only at integer filling factors (i.e., one electron or hole per moir\'e cell) but also at a series of fractional filling factors. The CIs at the integer filling factors are  driven primarily by the on-site repulsion in the Hubbard model, while those at factional filling factors are interpreted as generalized Wigner crystals \cite{regan2020mott,xu2020abundance,jin2020stripe,huang2020correlated} induced by the long-range Coulomb repulsion. The observed abundant correlated insulating states in twisted bilayer TMDs call for thorough theoretical investigations of this intriguing two-dimensional(2D) Moir\'e-Hubbard system.

In this Rapid Communication, we theoretically study a generalized Hubbard model on triangular moir\'e lattice realized in twisted bilayer TMDs. We show that the quantum phase diagram at a given fractional filling factor contains a rich set of competing phases that can be tuned by the twist angle $\theta$ and the dielectric environment. We also find that the phase diagram depends nontrvially on the filling factor. When interaction is much greater than the kinetic energy,  Wigner crystals generally form to minimize the long-range Coulomb interaction. The effective lattices of Wigner crystals depend sensitively on the filling factor, and can be triangle, rectangle, honeycomb, kagome, etc. After the electron spin degree of freedom is taken into account, spin models on distinct lattices can be simulated in this system by simply tuning the carrier density, leading to a variety of charge- and spin- ordered phases. In competition with these states derived from Wigner crystals, interaction-induced Chern insulators also appear in the phase diagram, which is remarkable since the non-interacting band structure in the model is topologically trivial. Here Chern insulators arise spontaneously from effective fluxes that are spontaneously generated either by nontrivial spin texture or by interaction-induced complex hopping phases. We elaborate our results by presenting calculated rich quantum phase diagrams at representative fractional filling factors, and discuss their experimental implications.

\begin{figure}[t]
	\centering
	\includegraphics[width=3.4in]{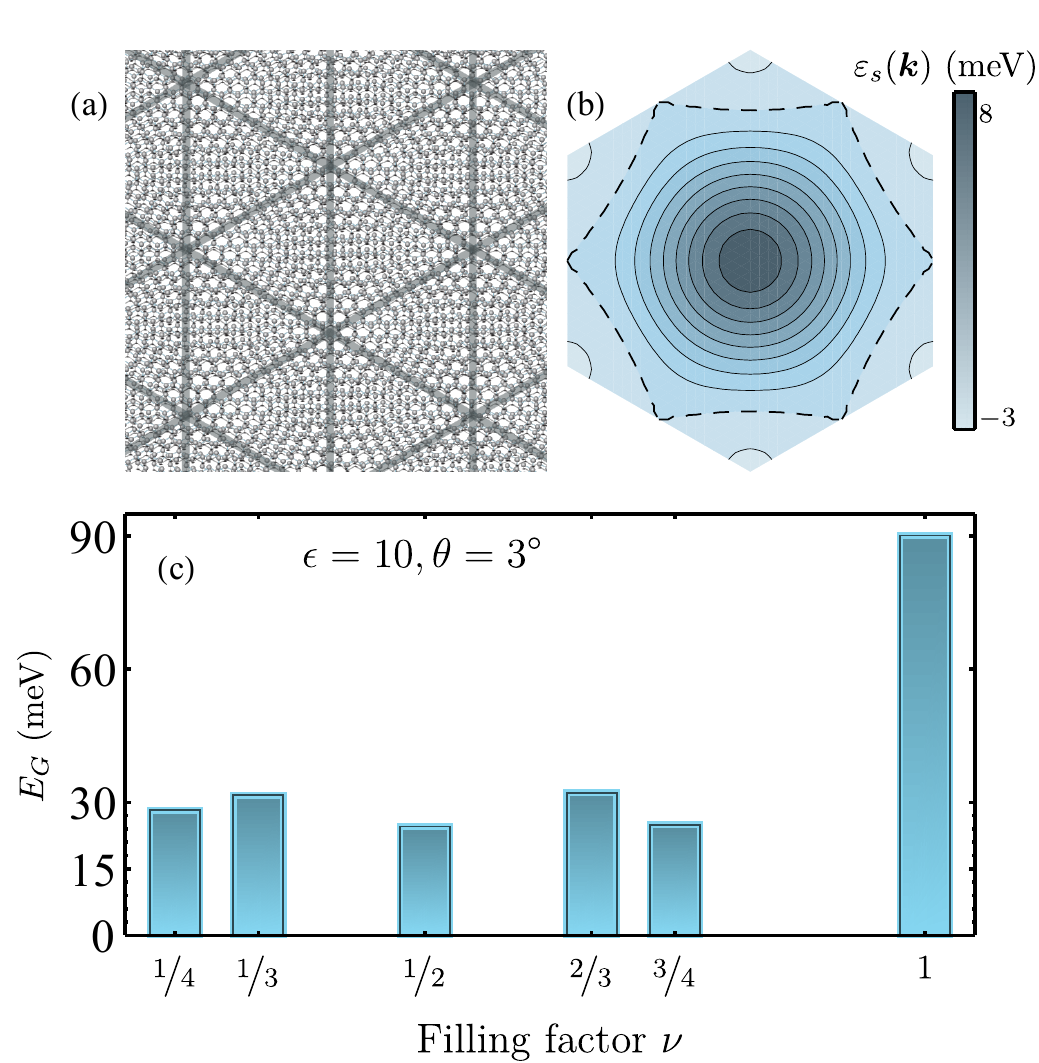}
	\caption{(a) The effective triangular lattice formed in the moir\'e pattern. (b) The single-particle moir\'e band $\varepsilon_s(\boldsymbol{k}) $ of Eq.~\eqref{eq:hubbard} at $ \theta=3^\circ $, where $s$ can be $\uparrow$ or $\downarrow$. The dashed line marks the contour at the  van Hove energy. (c) The correlated insulating gap at representative rational filling factors $ \nu $. }
	\label{fig:fig1}
\end{figure}

\textit{Model.}--- We study a moir\'e-Hubbard model defined as follows:
\begin{eqnarray}\label{eq:hubbard}
	H&=&\sum_{s}\sum_{i,j}^{} t_{s}\left(\bm{R}_i-\bm{R}_j\right) c_{i,s}^\dagger c_{j,s}\nonumber\\
	&+&\frac{1}{2}\sum_{s,s'}\sum_{i,j}U(\bm{R}_i-\bm{R}_j) c_{i,s}^\dagger c_{j,s'}^\dagger c_{j,s'} c_{i,s},
\end{eqnarray}
where $\bm{R}_i$ represents the position of site $i$ in a triangular lattice formed in the moir\'e pattern [Fig.~\ref{fig:fig1}(a)], $s$ is the spin index, and $t$ and $U$ are, respectively, the hopping parameter and the interaction strength. As proposed in Refs.~\cite{wu2018hubbard, wu2019topological, pan2020band}, the model in Eq.~\eqref{eq:hubbard} can be simulated in twisted TMD heterobilayers as well as homobilayers. For definiteness, we use twisted homobilayer WSe$_2$ (tWSe$_2$) as the model system in this work, and Eq.~\eqref{eq:hubbard} is then constructed following our previous work \cite{pan2020band} for low-energy holes in the first moir\'e valence band at $\pm K$ valleys. Here we use  $c_{i,s}^\dagger$ to represent the hole operator, and $s=\uparrow$ and $\downarrow$ are locked to $+K$ and $-K$ valleys, respectively. We define a filling factor $\nu$ as $(1/\mathcal{N}) \sum_{i,s} c_{i,s}^{\dagger} c_{i,s} $, which counts the number of holes per moir\'e cell ($\mathcal{N}$ is the total number of moir\'e sites in the system). The charge neutrality point of the semiconducting twisted bilayer corresponds to $\nu=0$. For simplicity, we assume that  no external out-of-plane displacement field is applied to WSe$_2$, and then the model in  Eq.~\eqref{eq:hubbard} respects emergent spin SU(2) symmetry and $C_6$ point group symmetry. An important advantage of the moir\'e platform is that both the hopping parameters and the interaction strength are highly tunable. Generally speaking, the moir\'e bandwidth becomes narrower at smaller twist angle (larger moir\'e period) and many-body interaction effects become more prominent~\cite{wu2018hubbard, wu2019topological, naik2018ultraflatbands}. We show the twist-angle dependence of $t$ and $U$ in the Supplemental Material \cite{SM} (see, also, Refs.~\cite{fukui2005chern,yu2011equivalent} therein). In the calculation of $U$, we project a screened Coulomb interaction $(e^2/\epsilon)(1/r-1/\sqrt{r^2+d^2})$ to the low-energy moir\'e states, where $\epsilon$ is the background dielectric constant that is tunable by the dielectric environment and $d/2$ is the distance between the moir\'e system and a nearby metallic gate. We take $\epsilon$ as a free parameter and  $d$, which is also experimentally controllable, to be 60 nm in calculations.  

We perform self-consistent mean-field (MF) Hartree-Fock studies of the moir\'e Hubbard model at representative filling factors with a variety of initial ansatze that range from  Wigner crystals (which can be derived from the classical Coulomb model~\cite{SM}) to topological states. At a given fractional filling factor, we generally find multiple solutions to the Hartree-Fock equation, and their energetic competitions give rise to  rich quantum phase diagrams. An overview of our results is illustrated in Fig.~\ref{fig:fig1}(c) showing the interaction-induced gap $E_G$ at rational $\nu$ with a denominator up to 4. In our theory, the CI at the integer filling $\nu=1$ is a Mott insulator, and its gap is primarily determined by the on-site repulsion~\cite{pan2020band}. CIs at fractional fillings often require the presence of off-site repulsion and generally have smaller charge gaps. The relative trend of our  calculated $ E_G $ in Fig.~\ref{fig:fig1} as a function of $\nu$ agrees well with a recent experiment in Ref.~\onlinecite{xu2020abundance}, which provides confidence in the validity of our theory.

\begin{figure}[t]
	\centering
	\includegraphics[width=3.4in]{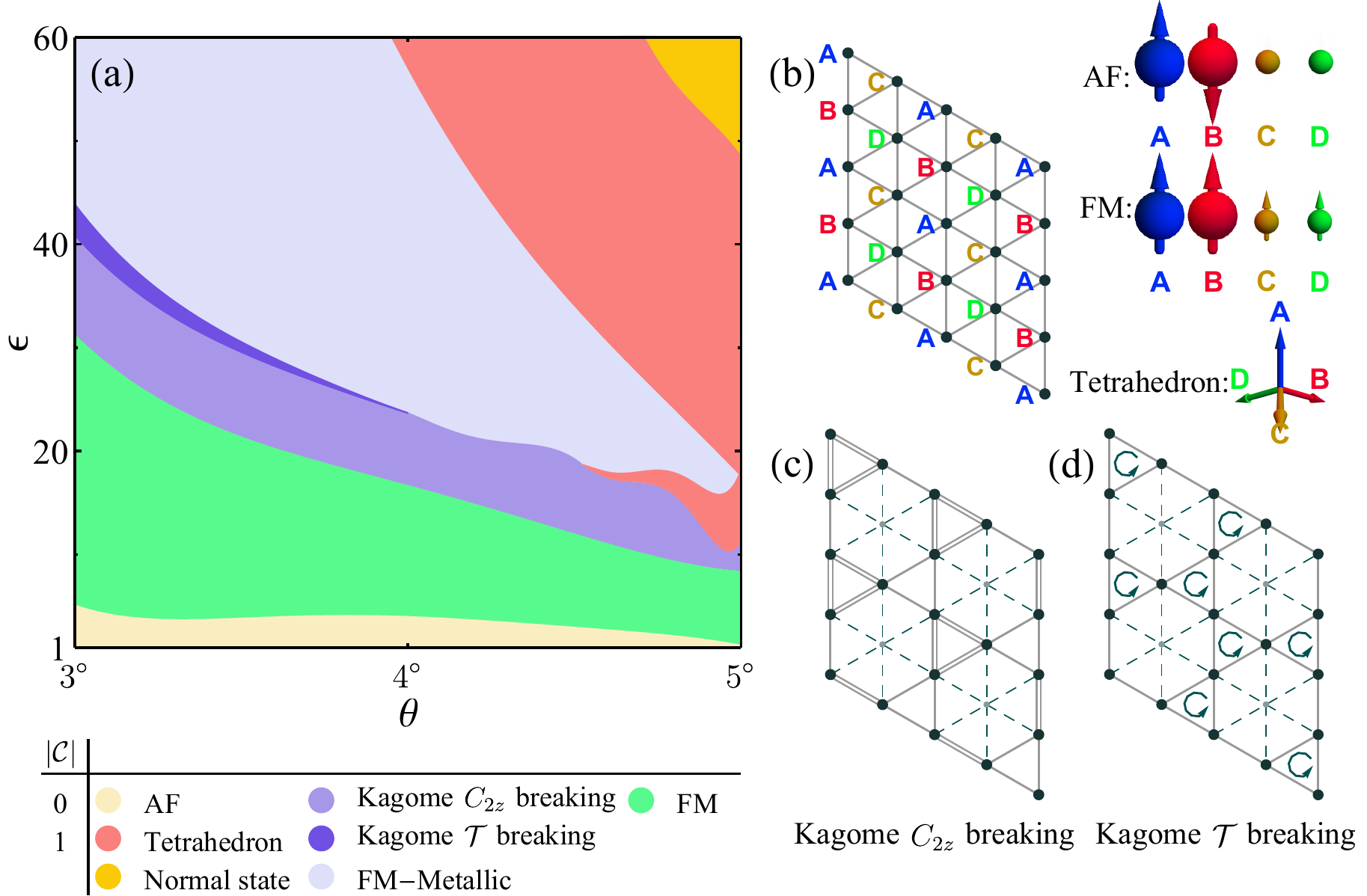}
	\caption{(a) The quantum phase diagram at $\nu=1/2$ as a function of $\theta$ and $\epsilon$. Some phases are illustrated in (b)-(d). (b) In the AF and FM phases, $A$ and $B$ sublattices are dominantly occupied, while $C$ and $D$ sublattices are less occupied. In the AF phase, spin polarization is antiparallel on $A$ and $B$, but vanishes on $C$ and $D$. In the FM phase, all sites have parallel spin polarization but different densities. In the tetrahedron phase, the four sublattices have equal density but different spin orientations that extend a solid angle of $4\pi$. (c) and (d)  show the kagome phases with $C_{2z}$ and $\mathcal{T}$ symmetry breaking, respectively.}
	\label{fig:fig2}
\end{figure}

$\nu=1/2$.--- The quantum phase diagram at $\nu=1/2$ is shown in Fig.~\ref{fig:fig2}(a), which displays six symmetry-breaking phases (besides a normal state without symmetry breaking) as a function of $\theta$ and $\epsilon$. When interaction is strong (small $\epsilon$), a Wigner crystal with a stripe charge density wave (CDW) forms  [Fig.~\ref{fig:fig2}(b)], and hosts a coupled-chain spin Heisenberg model. Our MF results show that the Heisenberg model has an antiferromagnetic (AF) exchange coupling, as an AF phase has a lower energy compared to the ferromagnetic (FM) phase for small $\epsilon$. When interaction decreases by increasing $\epsilon$, the stripe CDW gradually weakens and the FM phase becomes energetically more favorable. Therefore, charge and spin orderings are closely related. By further decreasing the interaction strength, CDW can completely disappear but the FM ordering can remain, which leads to a FM metallic phase. 

\begin{figure}[t]
	\centering
	\includegraphics[width=3.4in]{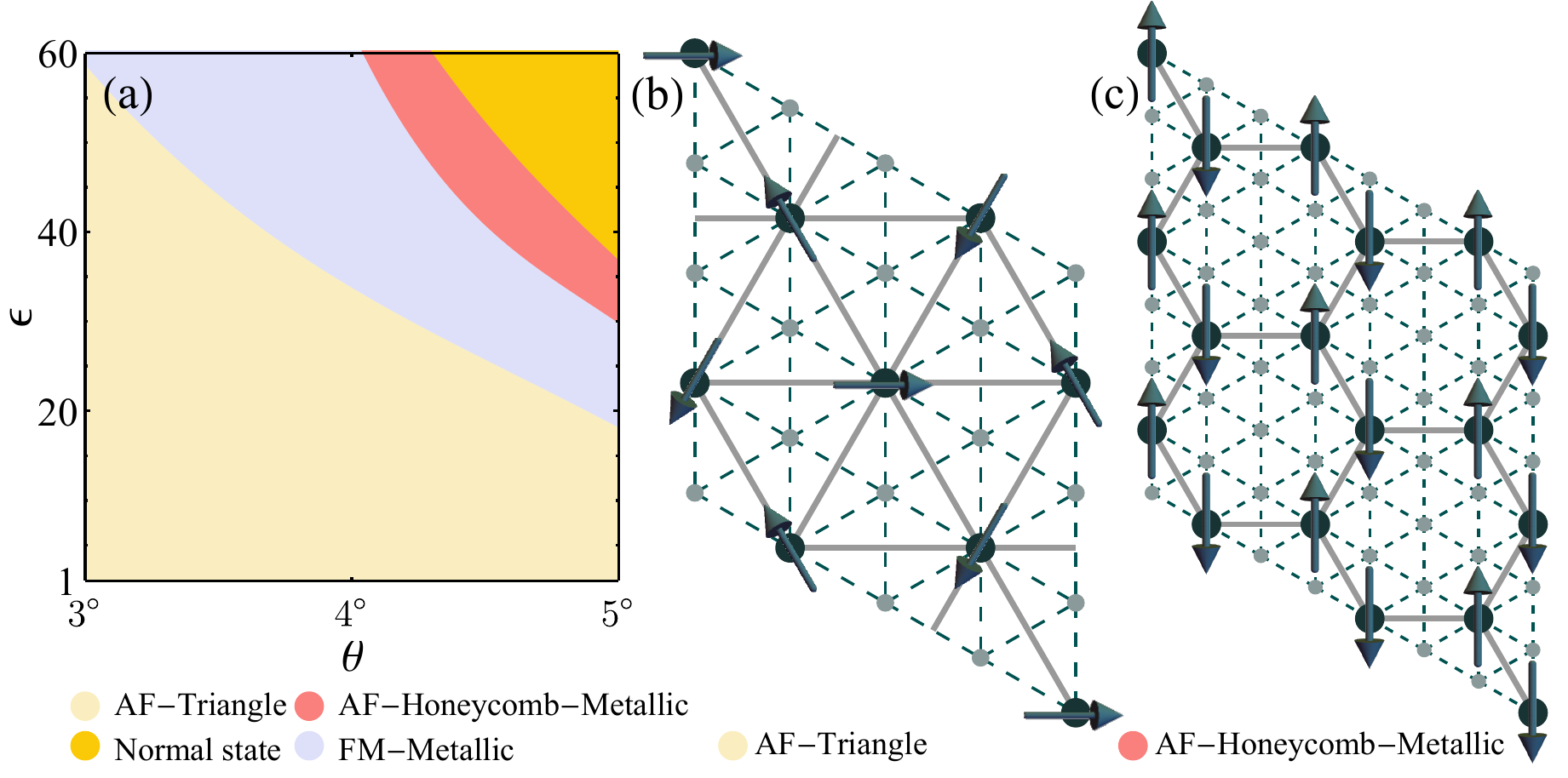}
	\caption{(a) The quantum phase diagram at $ \nu=1/3 $. Two of the phases are illustrated in (b) and (c).}
	\label{fig:pd1,3}
\end{figure}

In addition to these relatively simple charge- and spin-ordered phases, we also find three more exotic phases in Fig.~\ref{fig:fig2}(a): one tetrahedron phase and two kagome phases. In the tetrahedron phase, there is spin ordering but no charge ordering. The spin texture on the four magnetic sublattices forms a tetrahedron, which leads to a real-space Berry flux of $\pi$ for electronic motion along each triangular plaquette. We numerically verify that the tetrahedron phase is a Chern insulator with a Chern number of $|\mathcal{C}|=1$. This phase arises because our non-interacting moir\'e band at $\nu=1/2$ is close to the van Hove energy, and the corresponding Fermi surface is close to nesting [Fig.~\ref{fig:fig1}(b)], which leads to an instability towards noncollinear ordering \cite{martin2008itinerant}. In agreement with this weak-coupling picture, we find that the tetrahedron phase appears at relatively weak interactions.

In the two kagome phases shown in Figs.~\ref{fig:fig2}(c) and \ref{fig:fig2}(d), charge ordering leads to an effective kagome lattice where one out of four triangular sites are nearly unoccupied, and the other three sites each have a site occupancy $\sim 2/3$ and nearly full spin polarization. This spin- and charge-ordered kagome phase would host Dirac cones in the mean-field quasiparticle band structure at the Fermi energy, if there was no additional symmetry breaking. However, the Dirac cones can be gapped out by further breaking either twofold rotation $C_{2z}$ symmetry or time-reversal $\mathcal{T}$ symmetry. 

In the $C_{2z}$-breaking kagome phase, the interaction-renormalized effective hopping parameters from a site to its nearest neighbors on opposite directions become different but remain real [Fig.~\ref{fig:fig2}(c)], which leads to a valence bond solid insulator that is topologically trivial. In the other phase with $\mathcal{T}$ breaking, the effective hopping parameters acquire complex phases with a pattern shown in Fig.~\ref{fig:fig2}(d). This $\mathcal{T}$-breaking kagome phase with spontaneously-induced fluxes of $ \phi $ in the triangles and $ -2\phi $ in the hexagons is analogous to the Haldane model on honeycomb lattice \cite{haldane1988model}, and is a Chern insulator with $|\mathcal{C}|=1$~\cite{SM}. The topological kagome phase arising from a generalized Hubbard model on a triangular lattice has not been reported previously and provides a new mechanism to realize quantum anomalous Hall effect in a realistic experimental system.

\textit{$\nu=1/3$.}--- In the quantum phase diagram at $\nu=1/3$ shown in Fig.~\ref{fig:pd1,3}(a),  the Wigner crystal with a $\sqrt{3}\times \sqrt{3}$ CDW is robust up to very large $\epsilon$, and 120$^{\circ}$ AF order with a $3 \times 3$ period develops on top of this Wigner crystal [Fig.~\ref{fig:pd1,3}(b)]. For weak interactions, we find two metallic phases in addition to the normal state: (1) a FM metallic phase with spin polarization but no CDW; (2) an AF metallic phase [Fig.~\ref{fig:pd1,3}(c)] with a $3\times 3 $ CDW, where sites with dominant occupancy form an effective honeycomb lattice and host collinear AF ordering.

\textit{$\nu=2/3$.}--- The Wigner crystal at $\nu=2/3$ is dual to that at $\nu=1/3$, and forms a honeycomb lattice (Fig.~\ref{fig:pd2,3}), where spins develop collinear AF order in the strong interaction limit as expected from an effective Heisenberg model. By decreasing interaction, there is a transition from AF to FM spin orderings with the same $\sqrt{3} \times \sqrt{3} $ CDW, and then to FM without CDW, and finally to the normal state. We note that topological states derived from the Haldane model \cite{haldane1988model} can be Hartree-Fock solutions at both $\nu=1/3$ and $2/3$, but they are not energetically favorable within our explored parameter space~\cite{SM}.

\begin{figure}[t]
	\centering
	\includegraphics[width=3.4in]{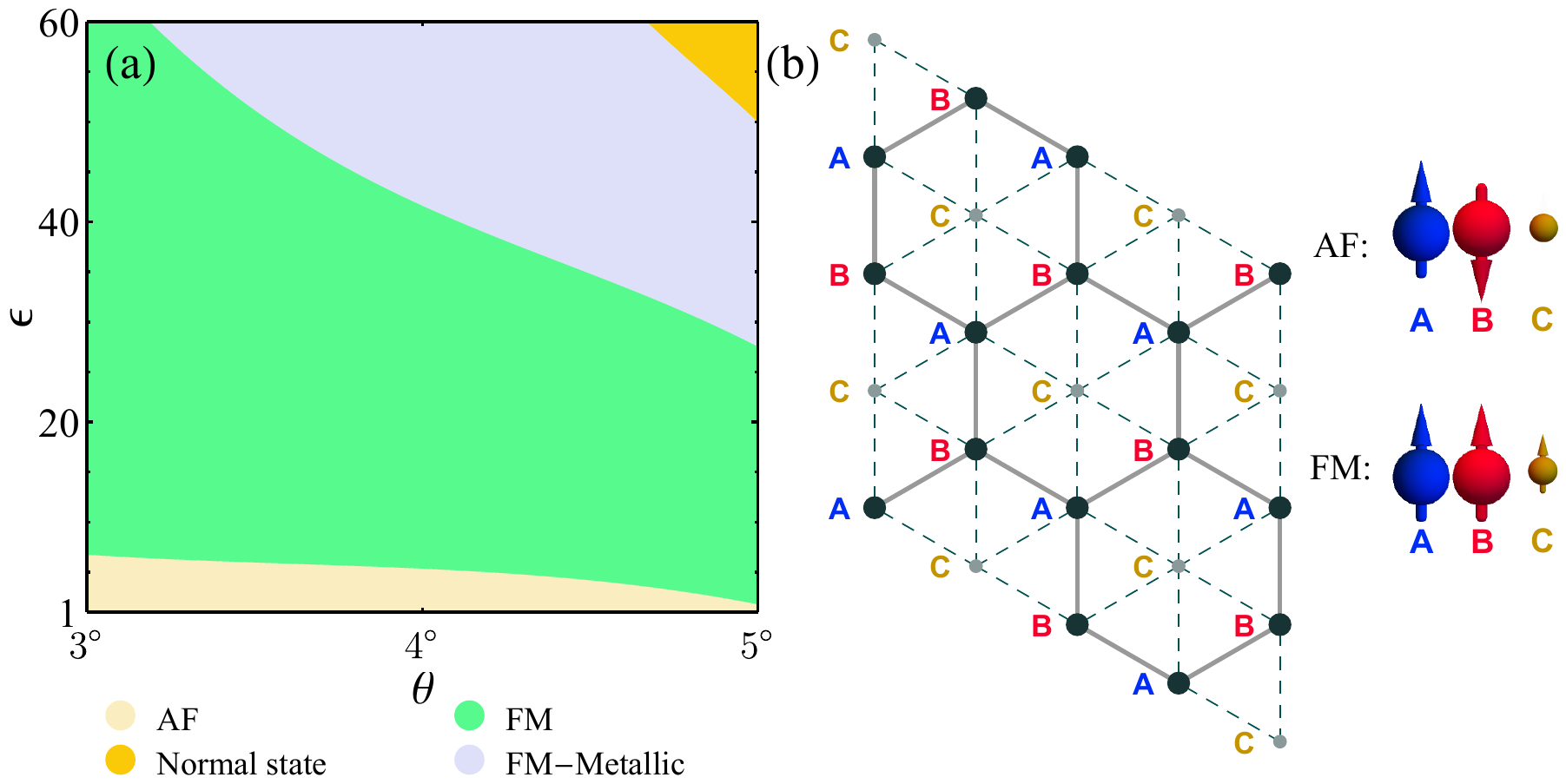}
	\caption{(a) The quantum phase diagram at $ \nu=2/3 $. (b)  AF and FM phases on an effective honeycomb lattice.}
	\label{fig:pd2,3}
\end{figure}

\textit{$\nu=1/4$.}--- At $\nu=1/4$, there are two types of Wigner crystals: (1) a $2\times 2$ triangular phase; and (2) a stripe phase with a $2\times \sqrt{3}$ rectangular superlattice, where the former appears in most of the parameter space in the phase diagram [ Fig.~\ref{fig:pd1,4}(a)] and the latter forms for small $\epsilon$ and large $\theta$. In both phases, the effective spin exchange interaction is weak because of the large separation (small hopping) between the primarily occupied sites, and therefore, AF and FM spin orderings closely compete in energy. We also find a Chern insulator state at $\nu=1/4$ that is analogous to the $\nu=1/2$ kagome phases with  $\mathcal{T}$ symmetry breaking, but it is energetically unfavorable~\cite{SM}.

\begin{figure}[t]
	\centering
	\includegraphics[width=3.4in]{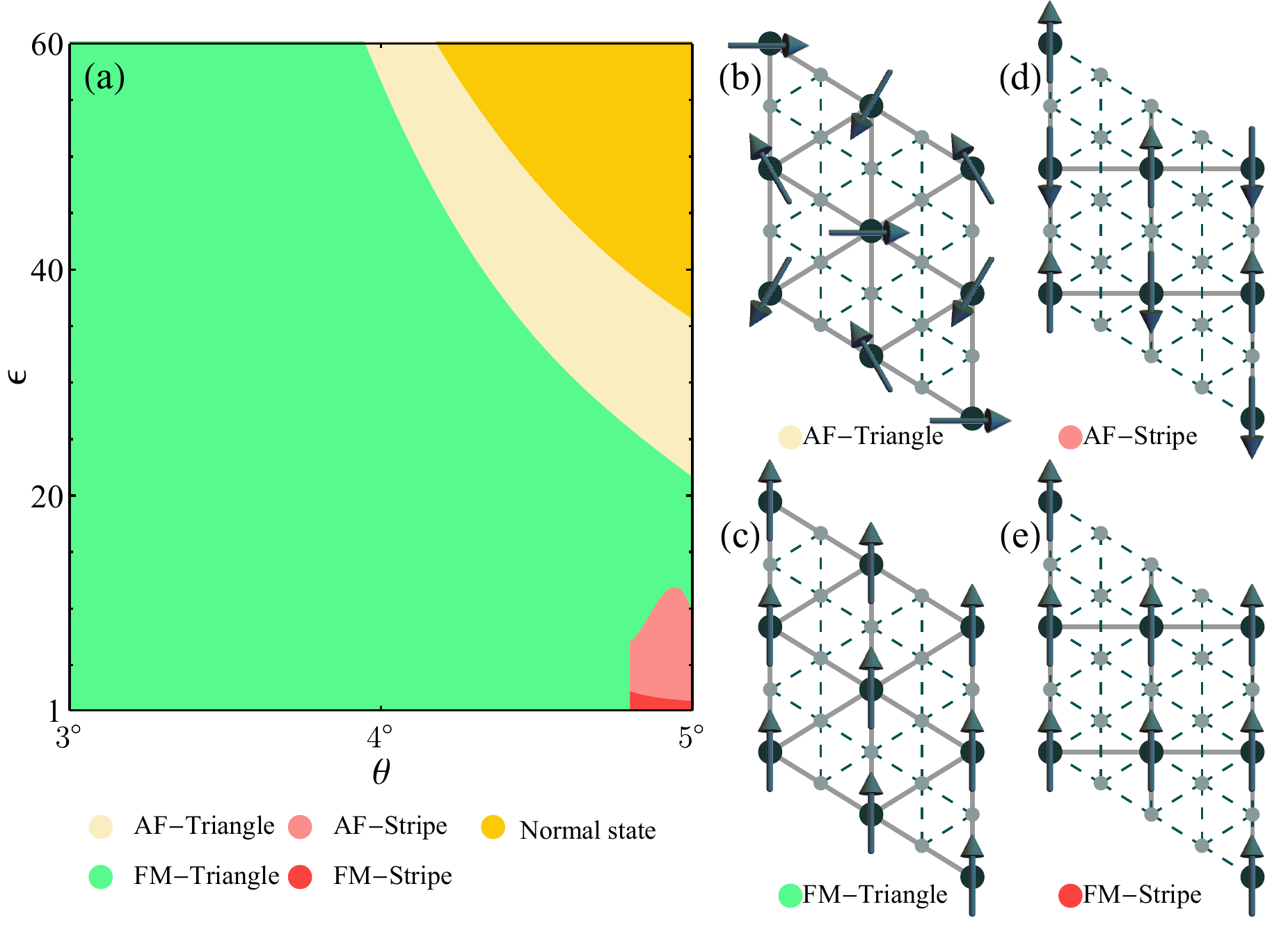}
	\caption{(a) The quantum phase diagram at $ \nu= 1/4 $. (b)  120$^{\circ}$ AF and (c) FM spin structures  on the $2\times 2$ triangular Wigner crystal. (d) Collinear AF and (e) FM spin structures on the $2 \times \sqrt{3} $ stripe Wigner crystal. }
	\label{fig:pd1,4}
\end{figure}

\textit{$\nu=3/4$.}---We find seven symmetry-breaking phases in the phase diagram at $ \nu=3/4 $, as shown in Fig.~\ref{fig:pd3,4}(a). For $
\epsilon <5$, we find two types of Wigner crystals, (1) a kagome lattice [Fig.~\ref{fig:pd3,4}(b)] for $\theta <4.2^{\circ}$ , and (2)  an anti-stripe lattice [Fig.~\ref{fig:pd3,4}(d)] for  $\theta >4.2^{\circ}$, which are, respectively, dual to the $2\times 2$ triangular and $2\times \sqrt{3}$ stripe Wigner crystals at $\nu=1/4$. We find that AF spin ordering has lower energy compared to FM spin ordering on both the kagome and anti-stripe lattices  for $ \epsilon <5$. It is important to note that both lattices with AF spin exchange couplings are frustrated and can host a large number of degenerate classical magnetic ground states, which could lead to quantum spin liquid states when quantum fluctuations in the spin sector are taken into account.

For  $	\epsilon > 5$, we find a FM phase on the kagome lattice , and the associated CDW gradually melts as $\epsilon$ increases, and finally vanishes, leading to a FM $1\times 1$ phase without CDW. In competition with this FM $1\times 1$ phase, there is a 120$^{\circ}$ AF phase that has only spin density wave but no CDW, as illustrated in Fig.~\ref{fig:pd3,4}(e).

Finally, we find two collinear AF phases that are derived from the kagome phases at $\nu=1/2$. Noting that $3/4=1/4+1/2$, we can construct collinear AF phases with effective filling factors of $1/4$ for the spin $\uparrow$ sector and $1/2$ for the spin $\downarrow$ sector. Spin $\uparrow$ and $\downarrow$ states, respectively, occupy sites on kagome and triangular lattices that are dual to each other. On the kagome lattice formed by spin $\downarrow$ states, $C_{2z}$ or $\mathcal{T}$ symmetry can be further broken, as in the case of $\nu=1/2$, leading to the two AF phases illustrated in Figs.~\ref{fig:pd3,4}(f) and \ref{fig:pd3,4}(g) that are respectively topologically trivial and nontrivial~\cite{SM}.

\begin{figure}[t]
	\centering
	\includegraphics[width=3.4in]{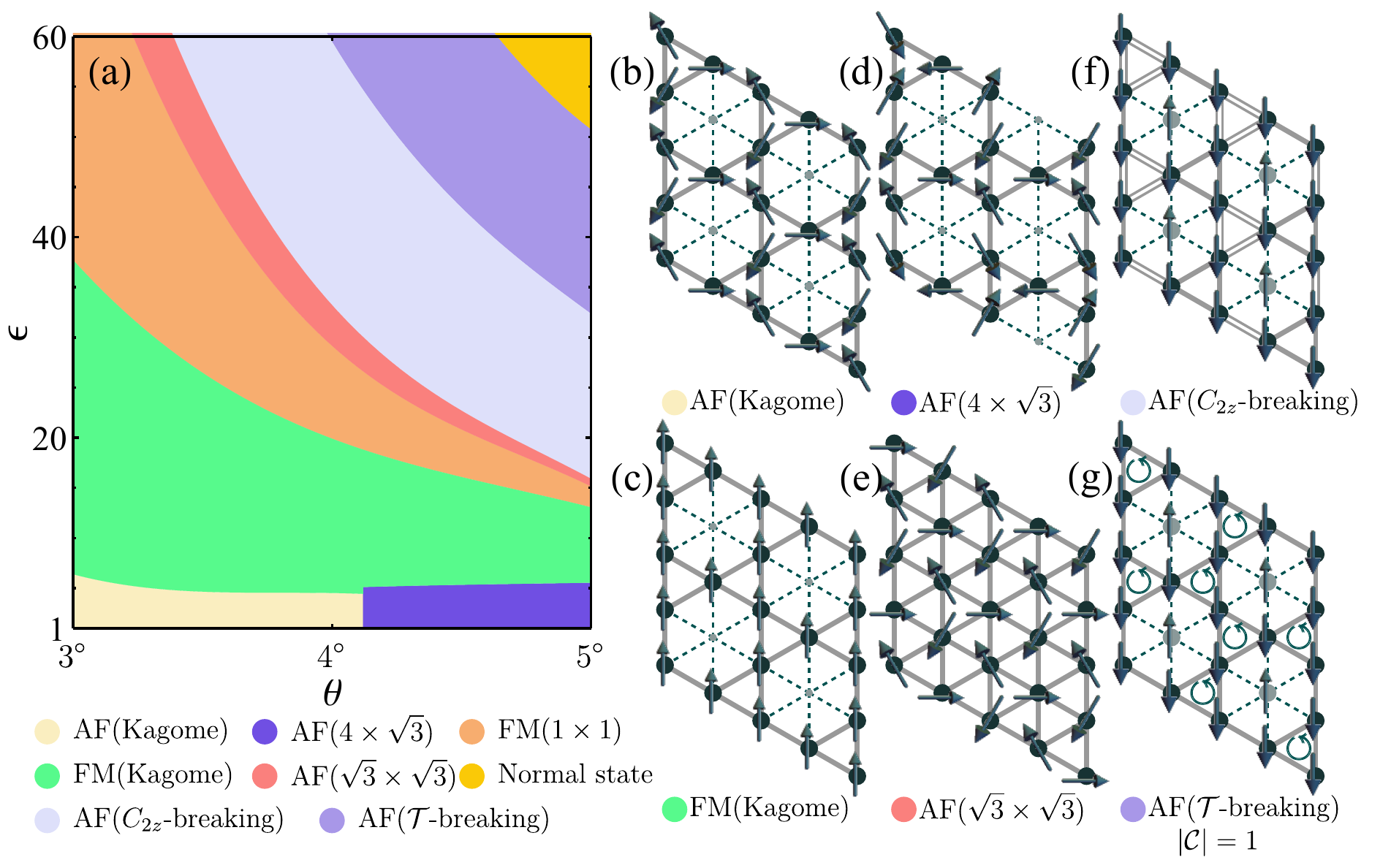}
	\caption{(a) The quantum phase diagram at  $ \nu=3/4 $. (b) AF and (c) FM spin structures on a kagome Wigner crystal. (d) AF spin structure on an anti-stripe Wigner crystal. The AF spin structures shown in (b) and (d) are mean-field results, and may not be the actual ground states because of fluctuations. (e) 120$^{\circ}$ AF state without charge density wave. (f) and (g) Sites with spin up (down) polarization form triangular (kagome) lattice. In (f), $C_{2z}$ symmetry is spontaneously broken, which leads to a valence bond solid insulator. In (g), $\mathcal{T}$ symmetry is spontaneously broken due to interaction-induced effective flux, which leads to a Chern insulator.}
	\label{fig:pd3,4}
\end{figure}

\textit{Discussions.}--- Our MF results should be taken to be qualitative instead of quantitative, as Hartree-Fock theory generally overestimates the tendency towards ordering. However, the advantage of MF theory is that it allows construction of a very large family of possible ground state candidate phases.  We envision that more sophisticated numerical approaches can be applied to the moir\'e Hubbard model, which could verify intriguing phases such as Chern insulators predicted by our theory and unveil more exotic phases, for example, spin liquid states on the effective kagome lattice at $\nu=3/4$, but such numerical methods are extremely computationally demanding and therefore, detailed results as functions of filling factors, interaction strength, and twist angle as provided in our work are challenging.  It is useful to mention here for comparison that the MF theory applied on the standard 2D minimal square-lattice on-site Hubbard model only finds a few phases (AF, FM, paramagnet, and spiral) as functions of interaction and filling~\cite{hirsch1985twodimensional,igoshev2015spiral}. Due to space limit, we only present phase diagrams at rational $\nu$ with a denominator up to 4, but we do also find correlated insulators at other fractional filling factors.

The predicted rich phase diagrams can lead to very rich experimental phenomena, because different phases can be accessed by tuning experimentally controllable parameters (e.g., $\theta$ and $\epsilon$).   Current experiments \cite{regan2020mott,tang2020simulation,wang2020correlated,jin2020stripe,xu2020abundance,huang2020correlated} were all performed using hexagonal boron nitride as encapsulating material. The corresponding dielectric constant $\epsilon$ is about 5$-$10.  For this range of $\epsilon$, our calculations show that ground states at the fractional filling factors are Wigner crystals. The effective lattice of Wigner crystals can spontaneously break threefold rotational symmetry, particularly in stripe phases at $\nu=1/2$ and $1/4$, which can be probed optically using linear dichroism~\cite{jin2020stripe}. To realize the predicted Chern insulators at $\nu=1/2$ and $3/4$, weaker interaction (i.e., $\epsilon>10$) is desirable, which can be engineered by changing the dielectric environment, for examples, using an encapsulating material with a higher dielectric constant and reducing the distance from the sample to the metallic gate. Experimental observation of such interaction-induced Chern insulators in a system with topologically trivial single-particle bands would greatly enhance the scope of quantum anomalous Hall effect.

\textit{Acknowledgments.} This work is supported by the Laboratory for Physical Sciences.

\bibliographystyle{apsrev4-1}
\bibliography{WC}

\clearpage
\setcounter{figure}{0}
\setcounter{equation}{0}
\renewcommand{\theequation}{S\arabic{equation}}
\renewcommand{\thefigure}{S\arabic{figure}}
\renewcommand{\thesection}{S\arabic{section}}

\onecolumngrid
\begin{center}
		\textbf{Supplemental Materials for ``Quantum Phase Diagram of a Moir\'e-Hubbard Model''}
\end{center}

\twocolumngrid

\section{Moir\'e Hamiltonian}
\begin{figure}[t]
	\centering
	\includegraphics[width=3.4in]{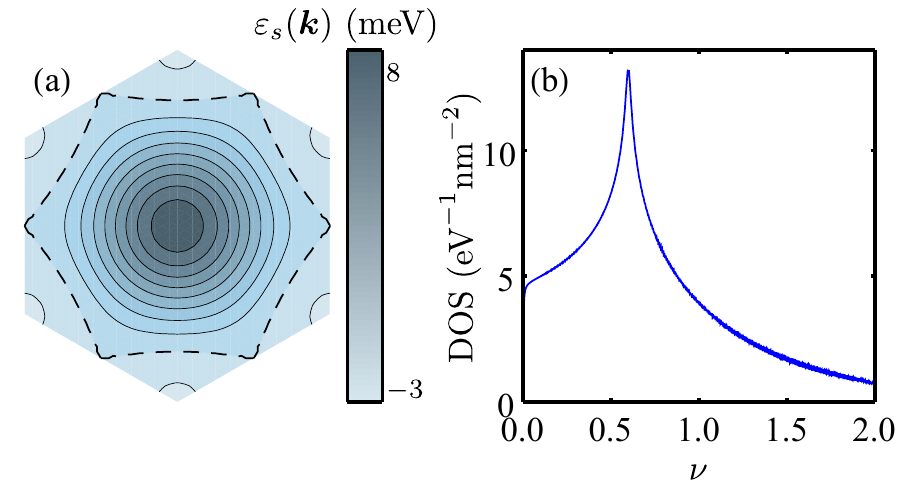}
	\caption{(a) The band structure and (b) density of states of the topmost valence moir\'e band at $ \theta=3^\circ $.}
	\label{fig:dos}
\end{figure}

The methodology to calculate moir\'e band structure for valence band states in twisted bilayer WSe$_2$ (tWSe$_2$) is given in Refs.~\onlinecite{wu2019topological,pan2020band}. Here, we briefly provide the nuermical details underlying our calculations. The moir\'e Hamiltonian for valence states in tWSe$_2$  at  $  +K $ valley is 
\begin{equation}\label{eq:Hmoire}
	\mathcal{H}_\uparrow=\begin{pmatrix}
		-\frac{\hbar^2 (\bm{k}-\bm{\kappa}_+)^2}{2m^*}+\Delta_{+}(\bm{r}) & \Delta_{\text{T}}(\bm{r})\\
		\Delta_{\text{T}}^\dagger(\bm{r}) & -\frac{\hbar^2(\bm{k}-\bm{\kappa}_-)^2}{2m^*}+\Delta_{-}(\bm{r})
	\end{pmatrix},
\end{equation}
where $ m^*=0.45m_0 $ is the valence band effective mass ($ m_0 $ is the rest mass of electron). The layer-dependent momentum offset $\bm{\kappa}_{\pm}=[4\pi/(3a_M)](-\sqrt{3}/2,\mp 1/2)$ capture the rotation in the momentum space, where $ a_M=a_0/\theta $ is the moir\'e lattice constant and $ a_0=3.28 $\AA{} is the lattice constant of monolayer WSe$ _{2} $. Here $\Delta_{\pm}(\bm{r})$ is the layer dependent moir\'e potential
\begin{equation}
	\Delta_{\pm}(\bm{r}) = 2 V \sum_{j=1,3,5}^{}\cos(\bm{b}_j\cdot \bm{r} \pm \psi),
\end{equation}
where $ \bm{b}_1=[4\pi/(\sqrt{3}a_M)](1,0)$ and $\bm{b}_j$ with $j=$2 to 6 are related to $\bm{b}_1$ by $(j-1)\pi/3$ rotation, and $ V $ and $ \psi $ characterize the amplitude and spatial pattern of the moir\'e potential. The interlayer tunneling $\Delta_{\text{T}}(\bm{r})$ is 
\begin{equation}
	\Delta_{\text{T}}(\bm{r}) = w (1+e^{-i \bm{b}_2 \cdot \bm{r}}+e^{-i \bm{b}_3 \cdot \bm{r}}),
\end{equation}
where $ w $ quantifies the interlayer tunneling strength. In this calculation, we choose a set of phenomenological parameters at which the topmost moir\'e valence band is topologically trivial: $ \qty(V,\psi,w)= $(4.4 meV, 5.9, 20 meV). 
We diagonalize the moir\'e Hamiltonian~\eqref{eq:Hmoire} using the plane-wave expansion based on Bloch's theorem, and then construct a generalized Hubbard model for the topologically trivial topmost valence band, which resides on an effective triangular lattice. To calculate the hopping energy $ t $ and Coulomb interaction $ U $ in the generalized Hubbard model, we first construct the Wannier function and choose the gauge which ensure the bottom-layer component of the Bloch wave function at each momentum to be real and positive at the origin in the real space. We shift the band structures to the vicinity of zero energy by dropping the onsite energy $ t_0 $, and flip the sign of the hopping parameters $t$ compared to those reported in Ref.~\cite{pan2020band}  since here our generalized Hubbard model is constructed using the hole operator while the moir\'e Hamiltonian~\eqref{eq:Hmoire} describes electron. The band structure and the density of states at $ \theta=3^\circ $ are shown in Fig.~\ref{fig:dos}, where the van Hove singularity is roughly at $ \nu=\frac{1}{2} $.

Figure~\ref{fig:tU} shows hopping $ t $  and Coulomb interaction $ U $ up to the first three neighbors, where the hopping $ t $ increases exponentially as twist angle $ \theta $ increases (moir\'e lattice constant $ a_M $ decreases) and $ U $ increases approximately linearly. In the numerical calculation, the three nearest neighbors are considered in the hopping term, while remote Coulomb interactions $ U $ up to 100 hexagonal shells are considered to guarantee the convergence. To calculate $U$, we project the screened Coulomb interaction $V(r)=(e^2/\epsilon)(1/r-1/\sqrt{r^2+d^2})$ onto Wannier states.  As shown in Fig.~\ref{fig:tU}, interaction $U(R)$ can be approximated by $V(R)$ for large $R$ as expected.

\section{Coulomb model}
The Wigner crystal as the ansatz for the Hubbard model is derived from a zero-temperature Coulomb model with only the potential term in the Hubbard model,
\begin{equation}
	H_{\text{Coulomb}}=\frac{1}{2}\sum_{s,s'}\sum_{i,j}U(\bm{R}_i-\bm{R}_j) n_{i,s}n_{j,s'},
\end{equation}
where $ n_{i,s}$ is the binary occupancy number of site $ i $. We choose proper supercells manually and minimize the total Coulomb energy per site by exploring various arrangement of occupied sites. 

Table ~\ref{tab:wc} lists all the possible Wigner crystals we find to be existing in the quantum phase diagram at different filling factors and also shows the minimal short-range interactions required to open a finite gap at such filling factors as well as the value of finite gap and energy per site correspondingly. While Table~\ref{tab:wc} presents analytical results for minimal interactions required for Wigner crystals,  our numerical calculations include interaction $U$ up to 100 hexagonal shells. 

\begin{table}[b]
	
	\caption{\label{tab:wc}Analytical results for Wigner crystals}
	\begin{ruledtabular}
		\begin{tabular}{lcccr}
			$ \nu $ & Wigner crystal & Least $ U_n $ & Gap & Energy\\
			\hline
			$ 1/2 $ & Stripe & $ U_0,U_1 $ &$ \min(U_0,2U_1) $ & $ \frac{U_1}{2} $\\
			$ 1/3 $ & Triangle & $ U_0,U_1 $& $ \min(U_0,3U_1) $ & $ 0 $\\
			$ 2/3 $ & Honeycomb & $ U_0,U_1 $ & $ \min(U_0,3U_1) $ & $ U_1 $\\
			$ 1/4 $ & Triangle & $ U_0,U_1 $ & $ \min(U_0,2U_1) $ & $ 0 $\\
			$ 1/4 $ & Stripe & $ U_0,U_1 $  & $ \min(U_0,2U_1) $ & $ 0 $\\
			$ 3/4 $ & Kagome  & $ U_0,U_1 $ & $ \min(U_0,2U_1) $ & $ \frac{3U_1}{2} $\\
			$ 3/4 $ & Anti-stripe & $ U_0,U_1 $ & $ \min(U_0-U_1,2U_1) $ & $ \frac{3U_1}{2} $ \\
			$ 1 $ & Mott insulator & $ U_0 $ & $ U_0 $ & $ 0 $
		\end{tabular}
	\end{ruledtabular}
\end{table}
\section{Hubbard model and mean-field theory}

\begin{figure}[t]
	\centering
	\includegraphics[width=3.4in]{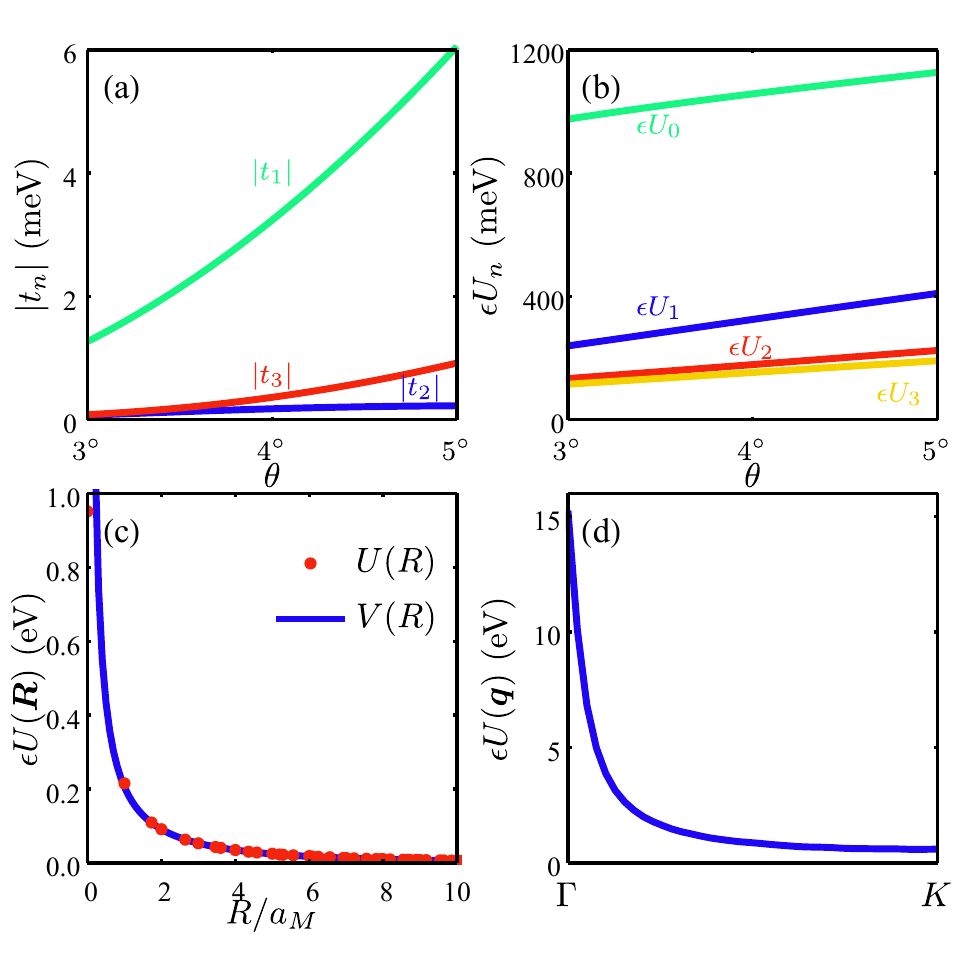}
	\caption{(a) and (b) $ \abs{t}_n $ and $\epsilon U_n $ as a function of twist angle $ \theta $. $ \epsilon $ is the effective dielectric constant. (c) $ U(R) $ can be approximated to $ V(R) $ with a good accuracy. Here $ \theta=3^\circ. $ (d) $ \epsilon U(\bm{q}) $ along $ \Gamma $ to $ K $ in one $ \mathbb{BZ} $ at  $ \theta=3^\circ$}
	\label{fig:tU}
\end{figure}

The Hubbard model of Eq.~(1) in the main text is composed of two terms: kinetic energy $ H_0 $ and interaction energy $ H_1 $. We perform the Fourier transformation of the Hamiltonian $ H $ in the real space to the momentum space. Therefore, the kinetic term $ H_0 $ becomes
\begin{equation}\label{eq:H0}
	H_0=\sum_{s}\sum_{{\bm{k}}}^{}\varepsilon_s(\bm{k}) c_{\bm{k},s}^\dagger c_{\bm{k},s},
\end{equation}
where $ \bm{k} $ is summed over the first Brillouin zone ($\mathbb{BZ}$) of the moir\'e lattice, and $ \varepsilon_s(\bm{k}) $ is the non-interacting band energy dispersion calculated from the tight-binding model for spin $ s $. The interaction term $ H_1 $ in the momentum space is 
\begin{equation}\label{eq:H1}
	H_1=\frac{1}{2\mathcal{N}}\sum_{s,s'} \sum_{\bm{k}} U(\bm{k}_\alpha-\bm{k}_\delta)\delta_{\bm{k}_\alpha,\bm{k}_\beta,\bm{k}_\gamma,\bm{k}_\delta} c_{\bm{k}_\alpha,s}^\dagger c_{\bm{k}_\beta,s'}^\dagger c_{\bm{k}_\gamma,s'} c_{\bm{k}_\delta,s},
\end{equation}
where $ \mathcal{N} $ is the number of total sites in the lattice, and  $ \bm{k}_\alpha,\bm{k}_\beta,\bm{k}_\gamma , \bm{k}_\delta $ are summed over the first $ \mathbb{BZ} $. Here, the interaction in the momentum space (as shown in Fig.~\ref{fig:tU}(d)) is
\begin{equation}
	U(\bm{q})=\sum_{\bm{R}}U(\bm{R})e^{i\bm{q}\cdot\bm{R}},
\end{equation}
and
\begin{equation}
	\delta_{\bm{k}_\alpha,\bm{k}_\beta,\bm{k}_\gamma,\bm{k}_\delta}=\sum_{\bm{G}}\delta(\bm{k}_\alpha+\bm{k}_\beta-\bm{k}_\gamma-\bm{k}_\delta,\bm{G}),
\end{equation}
where  $ \bm{G} $ is any moir\'e reciprocal lattice vector, and $ \delta(\dots) $ is the Kronecker delta function.

Using the Hartree-Fock truncation, we obtain the mean-field Hamiltonian for the interaction term
\begin{widetext}
	\begin{equation}\label{eq:HMF}
		H_{\text{int}}=\frac{1}{\mathcal{N}} \sum_{s,s'} \sum_{\bm{k}} U(\bm{k}_\alpha-\bm{k}_\delta) \delta_{\bm{k}_\alpha,\bm{k}_\beta,\bm{k}_\gamma,\bm{k}_\delta} \\
		\qty[\expval{c_{\bm{k}_\alpha,s}^\dagger c_{\bm{k}_\delta,s}}c_{\bm{k}_\beta,s'}^\dagger c_{\bm{k}_\gamma,s'}-\expval{c_{\bm{k}_\alpha,s}^\dagger c_{\bm{k}_\gamma,s'}}c_{\bm{k}_\beta,s'}^\dagger c_{\bm{k}_\delta,s}]
	\end{equation}
\end{widetext}

\begin{figure}[t]
	\centering
	\includegraphics[width=3.4in]{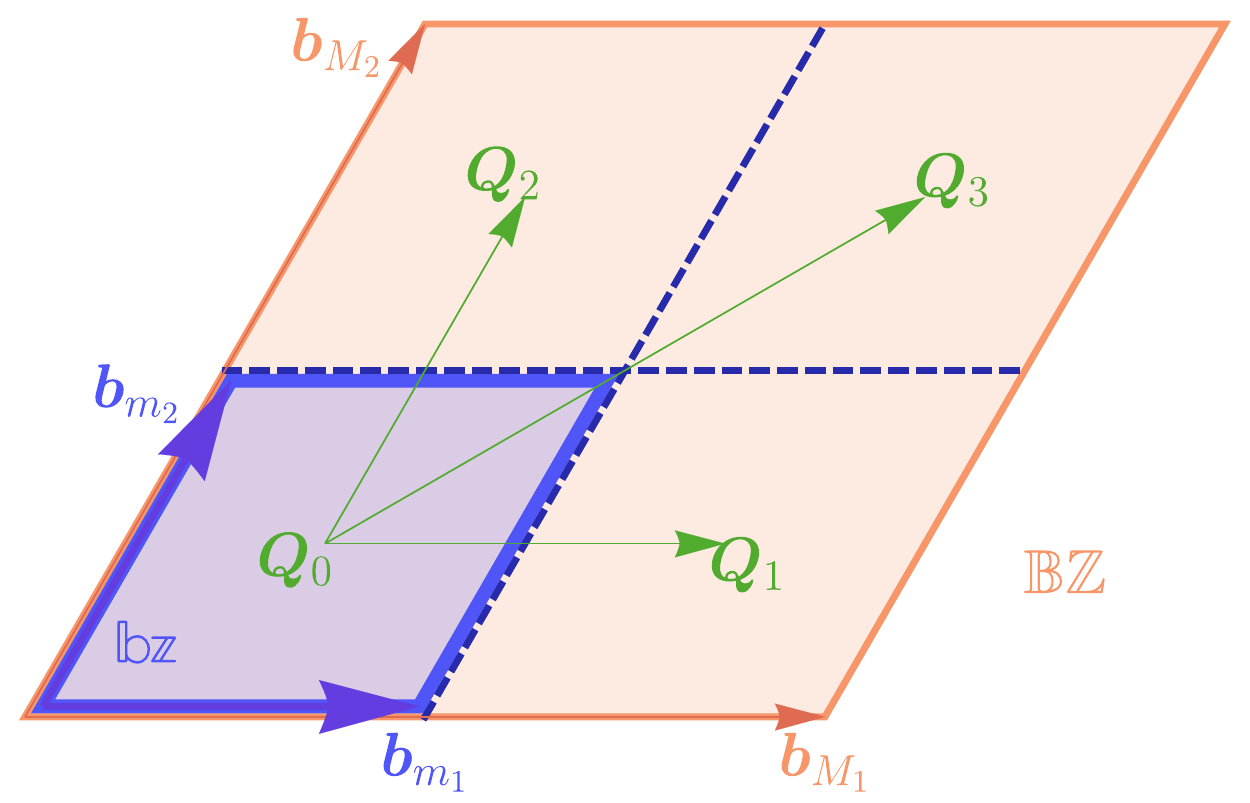}
	\caption{An example of $ \mathbb{bz} $ (blue) for the tetrahedron state at $ \nu=1/2 $ is quarter of the $ \mathbb{BZ} $ of moir\'e lattice (brown). The larger $ \mathbb{BZ} $ is tessellated by the smaller $ \mathbb{bz} $ with four shift vectors: $ \bm{Q}_0=\qty(0,0),\bm{Q}_1=\dfrac{\bm{b}_{M_1}}{2},\bm{Q}_2=\dfrac{\bm{b}_{M_2}}{2},\bm{Q}_3=\dfrac{\bm{b}_{M_1}+\bm{b}_{M_2}}{2} $.}
	\label{fig:bz}
\end{figure}

The Hartree-Fock state can spontaneously break the discrete translational symmetry, and resulting unit cell can be multiple times of the moir\'e unit cell, which causes the Brillouin zone ($ \mathbb{bz} $) to be smaller than the moir\'e Brillouin zone ($\mathbb{BZ}$). Therefore, $\mathbb{BZ}$ of moir\'e lattice can be tessellated by multiple $ \mathbb{bz} $s with appropriate shift vectors $ \bm{Q} $. (See Fig.~\ref{fig:bz} for example). Therefore, we can disassemble the summation over the whole $\mathbb{BZ}$ into aggregates  of several smaller  $ \mathbb{bz} $s with the shifting vectors, i.e., rewrite $ \bm{k}=\bm{q}+\bm{p} $, where $ \bm{q}\in \qty{\bm{Q}} $ and $ \bm{p} $ is a good quantum number lying in the smaller $ \mathbb{bz} $. Thus, Eq.~\eqref{eq:H0} becomes
\begin{equation}\label{eq:H0q}
	H_0= \sum_{s} \sum_{\bm{p},\bm{q}} \varepsilon_s(\bm{p}+\bm{q}) c_{\bm{p}+\bm{q},s}^\dagger c_{\bm{p}+\bm{q},s},
\end{equation}
where $ \qty{\bm{Q}} $ is the set of all shifting vectors, the number of $ \bm{Q} $ equals to the number of sites contained in one unit cell of the symmetry-breaking states.

Therefore, the mean-field Hamiltonian becomes
\begin{equation}
	H_{\text{HF}}= H_0+H_{\text{Hartree}}+H_{\text{Fock}},
\end{equation}
where the Hartree term is 
\begin{widetext}
	\begin{equation}\label{eq:hatree}
		H_{\text{Hartree}}=\frac{1}{\mathcal{N}} \sum_{s,s'} \sum_{{\bm{p},\bm{q}}}  U(\bm{q}_\alpha-\bm{q}_\delta) \delta_{\bm{q}_\alpha,\bm{q}_\beta,\bm{q}_\gamma,\bm{q}_\delta} \expval{c_{\bm{p}_\alpha+\bm{q}_\alpha,s}^\dagger c_{\bm{p}_\alpha+\bm{q}_\delta,s}} c_{\bm{p}_\beta+\bm{q}_\beta,s'}^\dagger c_{\bm{p}_\beta+\bm{q}_\gamma,s'}
	\end{equation}
\end{widetext}
and the Fork term is
\begin{widetext}
	\begin{equation}\label{eq:fock}
		H_{\text{Fock}}=-\frac{1}{\mathcal{N}}\sum_{s,s'} \sum_{{\bm{p},\bm{q}}} U(\bm{p}_\alpha-\bm{p}_\beta+\bm{q}_\alpha-\bm{q}_\delta)\delta_{\bm{q}_\alpha,\bm{q}_\beta,\bm{q}_\gamma,\bm{q}_\delta} \expval{c_{\bm{p}_\alpha+\bm{q}_\alpha,s}^\dagger c_{\bm{p}_\alpha+\bm{q}_\gamma,s'}} c_{\bm{p}_\beta+\bm{q}_\beta,s'}^\dagger c_{\bm{p}_\beta+\bm{q}_\delta,s}.
	\end{equation}	
\end{widetext}
Here the expected value $ \expval{\dots} $ is taken over all occupied states. We choose an initial ansatz for the Hartree-Fock state and substitute it into the $ H_{\text{HF}} $. After diagonalizating the $ H_{\text{HF}} $, we find the energies and wavefunctions, which are fed into the mean-field Hamiltonian $ H_{\text{HF}} $ again to find a self-consistent state iteratively. The convergence criterion is the total energy per site, which is defined as
\begin{widetext}
	\begin{eqnarray}\label{eq:totalE}
		\frac{\expval{H}}{\mathcal{N}}&=&\frac{1}{\mathcal{N}}\sum_{s} \sum_{\bm{p},\bm{q}}  \varepsilon_s\qty(\bm{p}+\bm{q})\expval{c_{\bm{p}+\bm{q},s}^\dagger c_{\bm{p}+\bm{q},s}}\\
		&+&\frac{1}{2\mathcal{N}^2}\sum_{s,s'}  \sum_{\bm{p},\bm{q}}   U(\bm{q}_\alpha-\bm{q}_\delta) \delta_{\bm{q}_\alpha,\bm{q}_\beta,\bm{q}_\gamma,\bm{q}_\delta} \expval{c_{\bm{p}_\alpha+\bm{q}_\alpha,s}^\dagger c_{\bm{p}_\alpha+\bm{q}_\delta,s}} \expval{c_{\bm{p}_\beta+\bm{q}_\beta,s'}^\dagger c_{\bm{p}_\beta+\bm{q}_\gamma,s'}}\\
		&-&\frac{1}{2\mathcal{N}^2}\sum_{s,s'}  \sum_{\bm{p},\bm{q}}  U(\bm{p}_\alpha-\bm{p}_\beta+\bm{q}_\alpha-\bm{q}_\delta)\delta_{\bm{q}_\alpha,\bm{q}_\beta,\bm{q}_\gamma,\bm{q}_\delta} \expval{c_{\bm{p}_\alpha+\bm{q}_\alpha,s}^\dagger c_{\bm{p}_\alpha+\bm{q}_\gamma,s'}} \expval{c_{\bm{p}_\beta+\bm{q}_\beta,s'}^\dagger c_{\bm{p}_\beta+\bm{q}_\delta,s}}.
	\end{eqnarray}
\end{widetext}

\section{Order parameter of the Wigner crystal}

\begin{figure}[t]
	\centering
	\includegraphics[width=3.4in]{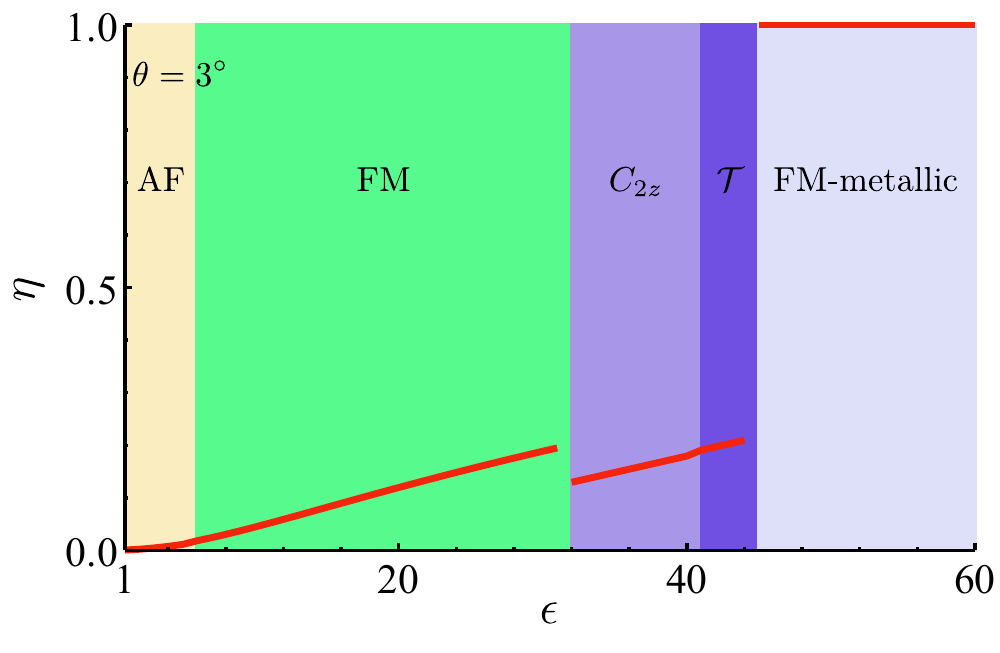}
	\caption{A line cut of order parameters of Wigner crystal at $ \theta=3^\circ $ and $ \nu=1/2 $.}
	\label{fig:minmax}
\end{figure}

We define the site-resolved density as:
\begin{equation}
	\expval{n_i}=\expval{{c}_{i,\uparrow}^\dagger {c}_{i,\uparrow}+{c}_{i,\downarrow}^\dagger {c}_{i,\downarrow}},
\end{equation}
where $ n $ is the average number density at site $ i $ in one unit cell. The order parameter of Wigner crystal is thus defined as
\begin{equation}\label{eq:eta}
	\eta=\frac{\min\limits_{i} n_i}{\max\limits_{i} n_i}.
\end{equation}
$ \eta\rightarrow0 $ indicates better Wigner crystallization while $ \eta=1 $ means no charge ordering in the moir\'e lattice. We present a line cut of order parameters $ \eta $ as a function of the background dielectric constant $ \epsilon $ at $ \theta=3^\circ $ and $ \nu=1/2 $ as shown in Fig.~\ref{fig:minmax}. The Wigner crystal is very well formed when $ \eta $ is small. At larger $ \epsilon $, $ \eta=1 $ indicates the Wigner crystal disappears in the FM-metallic phase--- each site is evenly occupied by half holes. The FM-metallic phase is thus a spin-polarized metal.

\section{The energy difference between AF and FM at $ \nu=1/4 $}
\begin{figure}[t]
	\centering
	\includegraphics[width=3.4in]{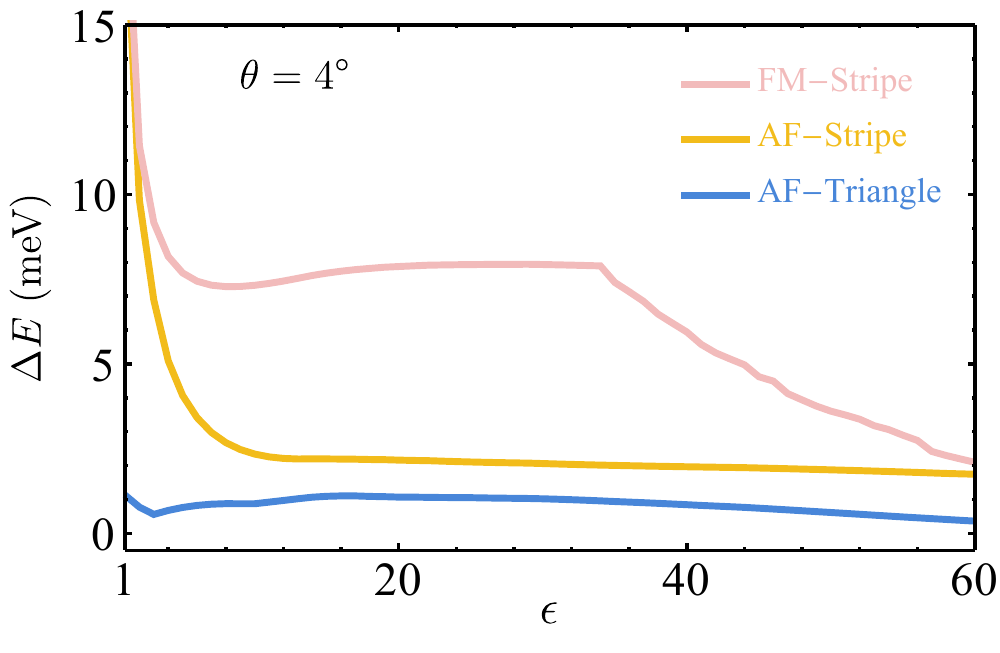}
	\caption{The energy per site of three phases--- AF-Triangle, AF-Stripe, and FM-Stripe--- relative to that of FM-triangle at $ \nu=\frac{1}{4} $ and $ \theta=4^\circ $.}
	\label{fig:diff}
\end{figure}

Figure~\ref{fig:diff} shows the energy of AF-Triangle (blue)/ AF-Stripe (yellow)/ FM-Stripe (orange) relative to that of FM-Triangle at $\nu=1/4$. In the main text, we find the competition of AF and FM is different in the phase diagram of $ \nu=\frac{1}{4} $ compared to other fractional $\nu$ , which we attribute to the larger site-to-site distance of the Wigner crystal and thus smaller exchange energy. Indeed, the energy of AF and FM in the phase diagram of $ \nu=1/4 $ is in a close competition. In Fig.~\ref{fig:diff}, we also find the energy of the triangle phase is smaller than that of the stripe phase.

\section{Effective Kagome lattice at $ \nu=1/2 $, $ 1/4 $, and $ 3/4 $: $ C_{2z} $ breaking vs $\mathcal{T}$ breaking}

\begin{figure}[t]
	\centering
	\includegraphics[width=3.4in]{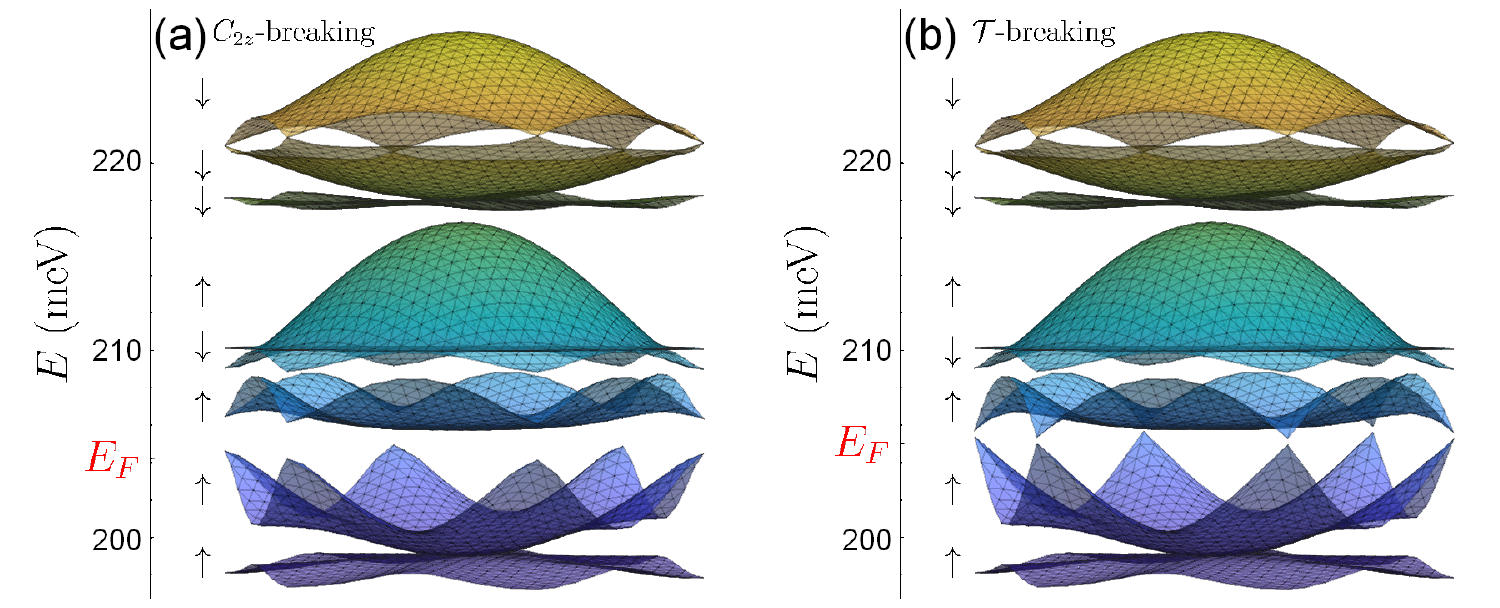}
	\caption{The interaction-renormalized band structure of the effective kagome lattice with (a) $ C_{2z} $-breaking and (b) $ \mathcal{T} $-breaking for $ \theta=3^\circ $ and $ \epsilon=35 $ at $ \nu=1/2 $. $\uparrow$ and $\downarrow$ label the spin polarization of each band, and $E_F$ indicates the Fermi energy.}
	\label{fig:kagome}
\end{figure}

Figure~\ref{fig:kagome}(a) and (b) show the interaction-renormalized band structure in the kagome phase due to the Coulomb repulsion at $ \nu=1/2 $, where the Dirac cone is opened at the corner of $ \mathbb{bz} $ due to the breaking of $ C_{2z} $ symmetry and $ \mathcal{T} $ symmetry respectively. The bottom two bands are occupied and fully polarized with spin $\uparrow$. The Fermi energy is labeled by $ E_F $.

To obtain these kagome phases, we introduce ancillary Hamiltonians. When $ C_{2z} $ symmetry breaks, we can construct spinless effective tight-binding model for valence bond solid insulator on the kagome lattice including only the nearest-neighbor hoppings,
\begin{equation}
	H_{C_{2z}}=\sum_{\expval{i,j}\in\qty{=}}t c_i^\dagger c_j+\sum_{\expval{i,j}\in\qty{-}}p c_i^\dagger c_j,
\end{equation}
where the nearest-neighbor pairs  $ \expval{i,j} $ are summed over single bonds $ \qty{-} $ with hopping $ p $ and double bonds $ \qty{=} $ with hopping $ t $ as shown in Fig.~2(c) in the main text,  The band structure can be obtained by transforming the ancillary Hamiltonian into the momentum space, i.e.,
\begin{widetext}
	\begin{equation}\label{eq:HC2}
		H_{C_{2z}}(\bm{k})= \mqty(0 & t e^{i \bm{k}\cdot \overrightarrow{AB}}+p e^{-i \bm{k}\cdot \overrightarrow{AB}} & t e^{i \bm{k}\cdot \overrightarrow{AC}}+ p e^{-i \bm{k}\cdot \overrightarrow{AC}}\\
		t e^{-i \bm{k}\cdot \overrightarrow{AB}}+p e^{i \bm{k}\cdot \overrightarrow{AB}} & 0 & t e^{i \bm{k}\cdot \overrightarrow{BC}}+p e^{-i \bm{k}\cdot \overrightarrow{BC}}\\
		t e^{-i \bm{k}\cdot \overrightarrow{AC}}+ p e^{i \bm{k}\cdot \overrightarrow{AC}} & t e^{-i \bm{k}\cdot \overrightarrow{BC}}+p e^{i \bm{k}\cdot \overrightarrow{BC}} & 0),
	\end{equation}
\end{widetext}

where site $ A,B,C $ are defined in Fig.~2(a), and $ \overrightarrow{AB},\overrightarrow{AC},\overrightarrow{BC} $ are all defined on the double bond plaquette as shown in Fig.~2(c) in the main text. We diagonalize the ancillary Hamiltonian~\eqref{eq:HC2} to obtain the wavefunction, which will be used as the initial ansatz before the first iteration.

\begin{figure}[t]
	\centering
	\includegraphics[width=3.4in]{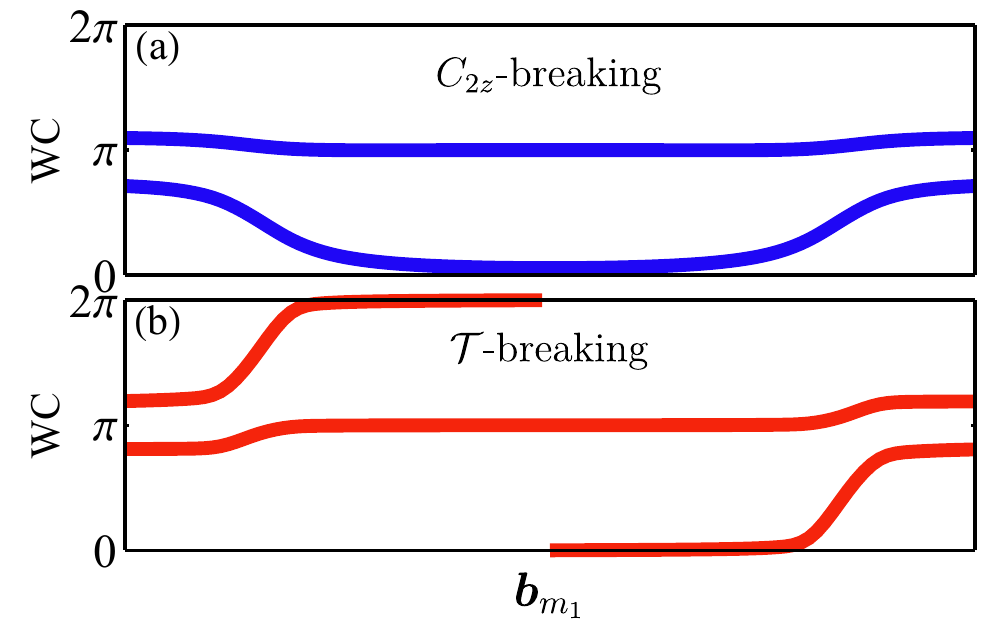}
	\caption{Wannier center flow along $ \bm{b}_{m_1} $ for the topologically trivial $ C_2 $-breaking (a) and topologically nontrivial $ \mathcal{T} $-breaking (b) for $ \theta=3^\circ $ and $ \epsilon=35 $ at $ \nu=1/2 $.}
	\label{fig:WCF}
\end{figure}

We calculate the Chern number of all the occupied bands~\cite{fukui2005chern} in Fig.~\ref{fig:kagome}(a) and find $ \abs{\mathcal{C}}=0 $. We also show Wannier center (WC) flow~\cite{yu2011equivalent} along one reciprocal vector in Fig.~\ref{fig:WCF}, which also has zero winding. The Wannier center is defined here as the phases of eigenvalues of a Wilson loop along a closed path $ L $, i.e., $ \arg(\exp(i{\oint_L A(k)dk})), $
where $ A(k) $ is the non-Abelian berry connection. Here we choose the closed path $ L $ along $ \bm{b}_{m_2} $ and plot the Wannier center flow along the direction of $ \bm{b}_{m_1} $.


When $ \mathcal{T} $ symmetry breaks, we can construct a spinless model for the Chern insulator by imposing complex hoppings on the nearest neighbors,
\begin{equation}\label{eq:HT}
	H_{\mathcal{T}}=\sum_{\expval{i,j}} t e^{i\phi/3 v_{i,j}} c_i^\dagger c_j,
\end{equation}
where $ v_{i,j}= $ 1 (-1) if the hopping from $ j $ to $ i $ is counterclockwise (clockwise) in the triangular plaquette in the kagome lattice, and $ \phi $ is the nonzero effective flux. To find the band structure in the momentum space, we perform the Fourier transformation and obtain
\begin{widetext}
	\begin{equation}\label{eq:HTk}
		H_{\mathcal{T}}(\bm{k})=\mqty(0 & t e^{i\phi/3}\cos(\bm{k}\cdot\overrightarrow{AB}) & t e^{-i\phi/3}\cos(\bm{k}\cdot\overrightarrow{AC})\\
		t e^{-i\phi/3}\cos(\bm{k}\cdot\overrightarrow{AB}) & 0 & t e^{i\phi/3}\cos(\bm{k}\cdot\overrightarrow{BC})\\
		t e^{i\phi/3}\cos(\bm{k}\cdot\overrightarrow{AC}) & t e^{-i\phi/3}\cos(\bm{k}\cdot\overrightarrow{BC}) & 0
		),
	\end{equation}
\end{widetext}

where site $ A,B,C $ are defined in Fig.~2(a) in the main text, and $ \overrightarrow{AB},\overrightarrow{AC},\overrightarrow{BC} $ are defined on the triangles pointing to the right in Fig.~2(d) in the main text. The Dirac cones at $ \mathbb{bz} $ corners are gapped out as long as $ \phi\neq n\pi $, where $ n\in\mathbb{Z} $. Therefore, we choose $ \phi=\pi/2 $ and diagonalize the ancillary Hamiltonian~\eqref{eq:HTk}to obtain its wavefunction as the initial ansatz before the first iteration of the Hartree-Fock calculation. This leads to an intrinsic zero-magnetic-field quantum Hall effect, a kagome analog of Haldane model, which is topologically nontrivial and the Wannier center flow winds one time along the reciprocal unit vector $ \bm{b}_{m_1} $ as shown in Fig.~\ref{fig:WCF}(b). 

\begin{figure}[t]
	\centering
	\includegraphics[width=3.4in]{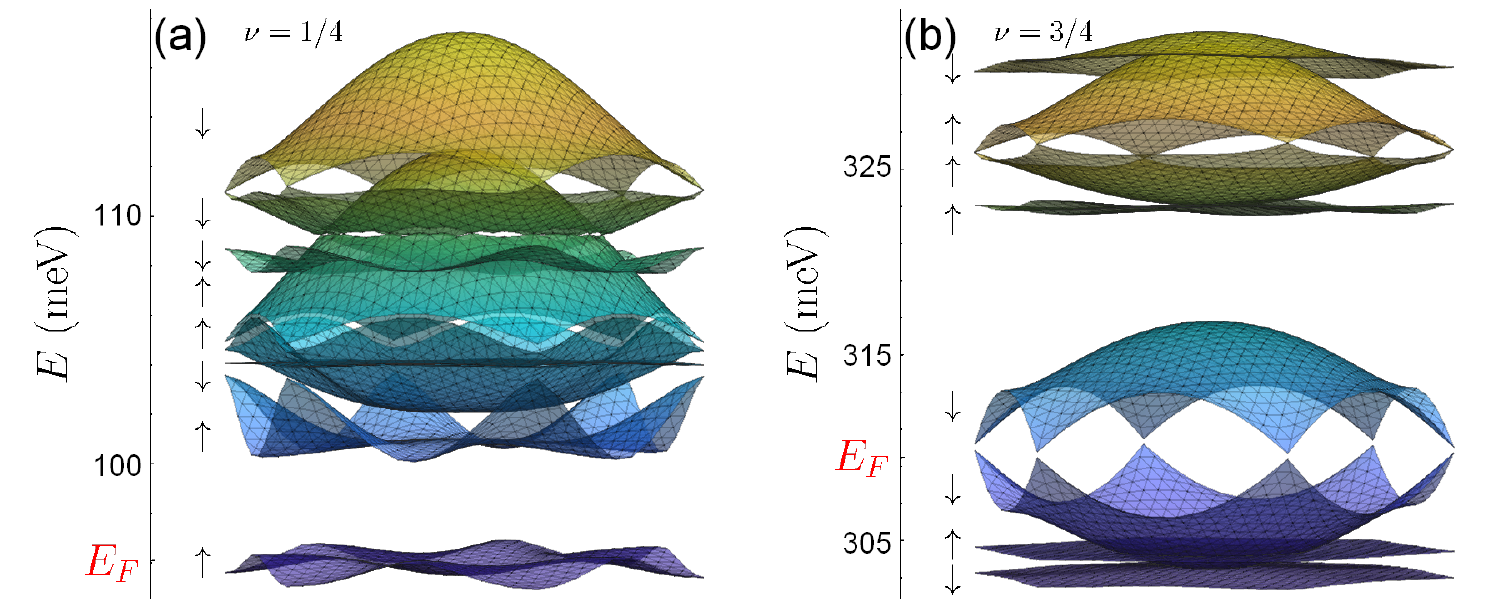}
	\caption{The interaction-renormalized band structure of the effective kagome lattice with $ \mathcal{T} $ breaking at (a) $ \nu=1/4 $ and (b) $ \nu=3/4 $ for $ \theta=3^\circ $ and $ \epsilon=35 $.}
	\label{fig:kagome4}
\end{figure}

\begin{figure}[t]
	\centering
	\includegraphics[width=3.4in]{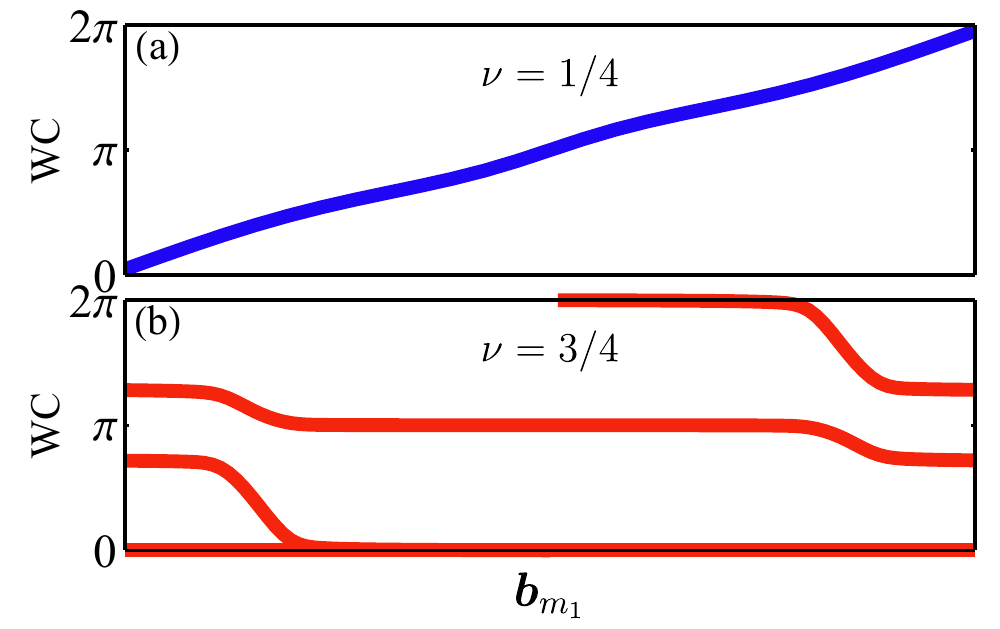}
	\caption{Wannier center flow along $ \bm{b}_{m_1} $ for the $ \mathcal{T} $ symmetry breaking case at (a) $ \nu=1/4 $ and (b) $ \nu=3/4 $ for $ \theta=3^\circ $ and $ \epsilon=35 $.}
	\label{fig:WCF4}
\end{figure}

Similarly, at $ \nu=1/4 $, there are also two kinds of kagome lattice with $ C_{2z} $ symmetry breaking and $ \mathcal{T} $ symmetry breaking. Figure~\ref{fig:kagome4}(a) shows the interaction-renormalized band structures for the $ \mathcal{T} $ symmetry breaking and Fig.~\ref{fig:WCF4}(a) shows its topologically nontrivial Wannier center flow at $ \nu=1/4 $. These kagome phases at $\nu=1/4$ are meta-stable states that are energetically unfavorable.

In the $\nu=3/4$ topological phase illustrated in Fig. 6(g) of the main text, spin $\downarrow$ states occupy a kagome lattice, while spin $\uparrow$ states occupy a triangular lattice. The corresponding band structure is shown  in Fig.~\ref{fig:kagome4}(b), where band structures derived from kagome (spin $\downarrow$) and triangular (spin $\uparrow$) lattices can be identified. The Wannier center flow shown in Fig. 9(b) confirms that the state has a Chern number of $|\mathcal{C}|=1$.

\section{Effective Honeycomb lattice at $ \nu=1/3 $ and $ 2/3 $: Haldane model}

\begin{figure}[t]
	\centering
	\includegraphics[width=3.4in]{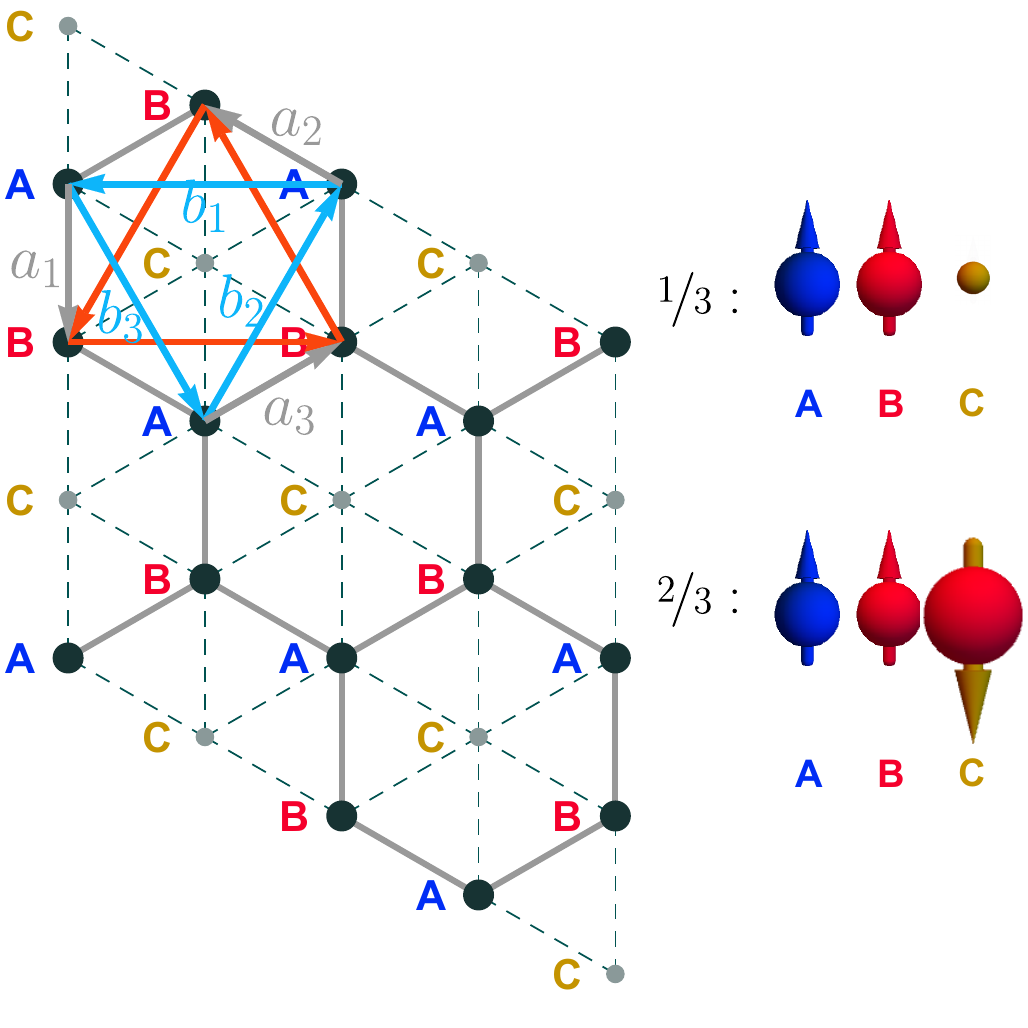}
	\caption{Two Topological states at $ \nu=1/3 $ and $ \nu=2/3 $ that are constructed based on the Haldane model. The hoppings between next-nearest neighbors (red and blue) are complex.}
	\label{fig:haldane}
\end{figure}

At $ \nu=1/3 $ and $ \nu=2/3 $, we can also construct topologically nontrivial states, although we find them to be energetically unfavorable. We derive the initial ansatz from the Haldane model~\cite{haldane1988model,raghu2008topological} by introducing a nonzero phase on the next-nearest-neighbors of the honeycomb as shown in Fig.~\ref{fig:haldane}. For example at $ \nu=1/3 $, the hoppings between the neighboring sites $ A $ ($ B $) following the blue (red) arrows are $ {t_2} e^{i\phi} $ ($ t_2 $ is real); the hoppings between the nearest $ A $ and $ B $ sites are the real $ t_1  $. The corresponding ancillary Hamiltonian in the momentum space is
\begin{equation}\label{eq:hhc}
	H_{\text{HC}}(\bm{k})=\mqty(2t_2 \sum\limits_{\bm{b}} \cos(\bm{k} \cdot \bm{b}-\phi) & t_1 \sum\limits_{\bm{a}} e^{i \bm{k}\cdot \bm{a}}\\
	t_1 \sum\limits_{\bm{a}} e^{-i \bm{k}\cdot \bm{a}}  & 2t_2 \sum\limits_{\bm{b}} \cos(\bm{k}\cdot \bm{b}+\phi)
	),
\end{equation}
where three $ \bm{a} $ connect the three pairs of the nearest-neighbors $ \overrightarrow{AB} $, and three $ \bm{b} $ connecting the next-nearest neighbors are defined as $ \bm{b}_1=\bm{a}_2-\bm{a}_3 $, $ \bm{b}_2=\bm{a}_3-\bm{a}_1 $, and $ \bm{b}_3=\bm{a}_1-\bm{a}_2 $. 

\begin{figure}[t]
	\centering
	\includegraphics[width=3.4in]{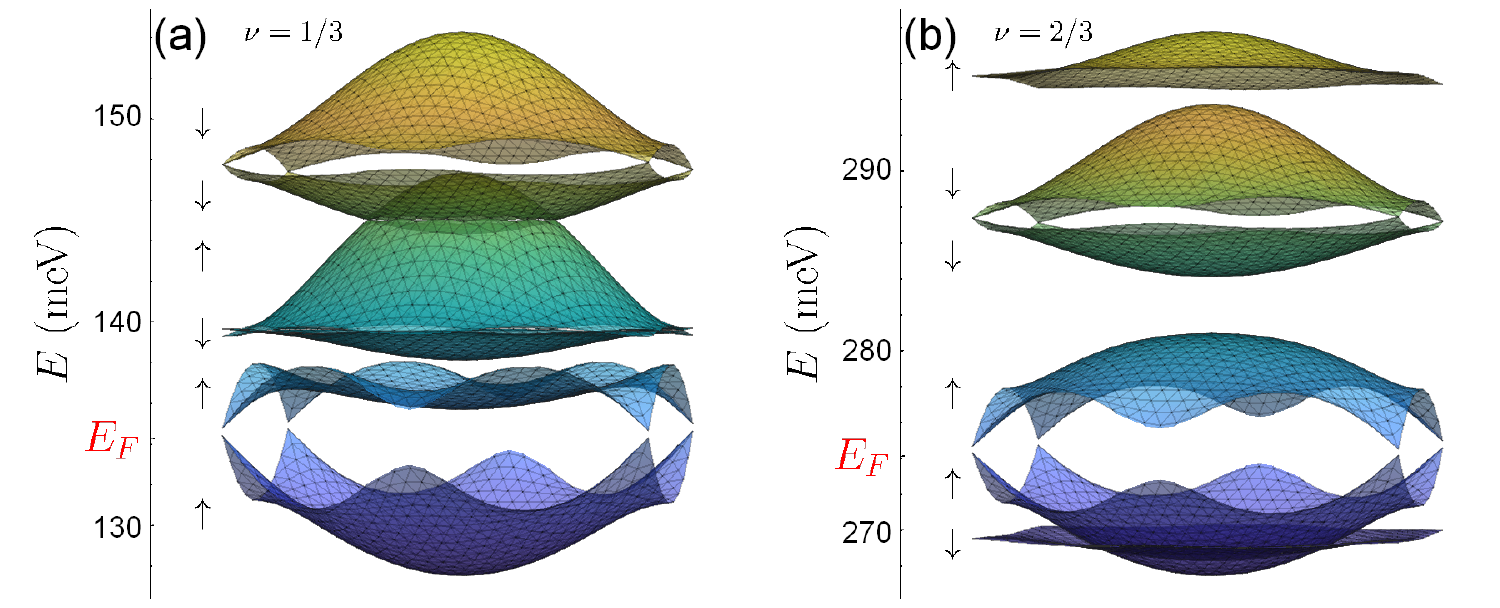}
	\caption{The interaction-renormalized band structure of the effective kagome lattice with $ \mathcal{T} $ breaking at (a) $ \nu=1/3 $ and (b) $ \nu=2/3 $ for $ \theta=3^\circ $ and $ \epsilon=35 $.}
	\label{fig:honeycomb3}
\end{figure}

\begin{figure}[t]
	\centering
	\includegraphics[width=3.4in]{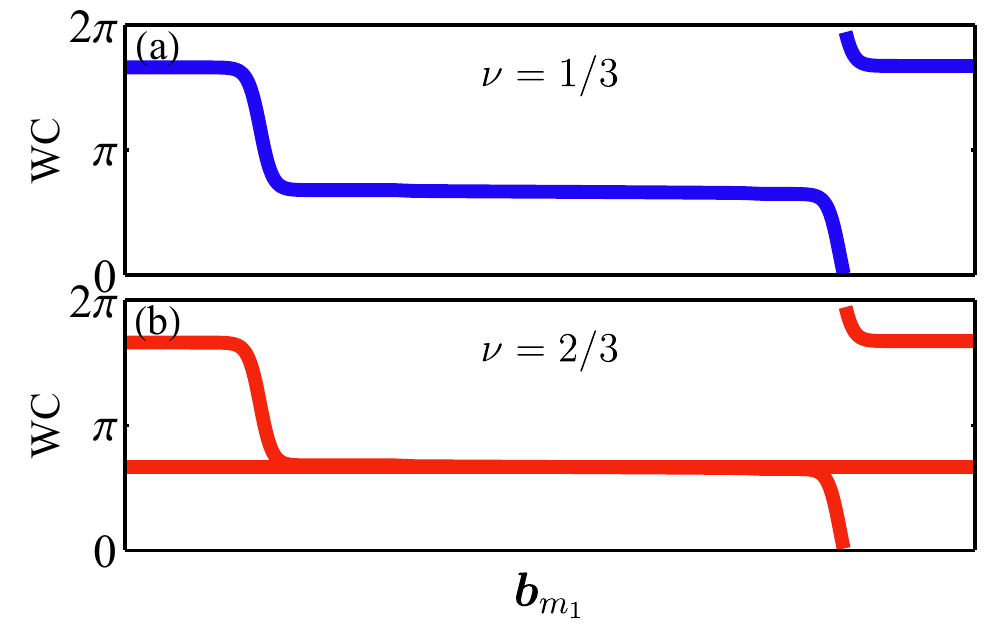}
	\caption{Wannier center flow along $ \bm{b}_{m_1} $ for topological states at (a) $ \nu=1/3 $ and (b) $ \nu=2/3 $ for $ \theta=3^\circ $ and $ \epsilon=35 $.}
	\label{fig:WCF3}
\end{figure}
We diagonalize Eq.~\eqref{eq:hhc} and use wavefunction as the initial ansatz of the Hubbard model. Figures~\ref{fig:honeycomb3}(a) and~\ref{fig:WCF3}(a) show the interaction-renormalized bandstructures and the corresponding nontrivial Wannier center flow at $ \nu=1/3 $.

At $ \nu=2/3 $, we can also construct a topological state inspired by $ 2/3=1/3+1/3 $, where sites $ A,B $ host spin $\uparrow$ with half occupancy and site $ C $ hosts spin $\downarrow$ with unity occupancy. Therefore, sites $ A,B $ form a  honeycomb lattice of Haldane model and sites $ C $ form a triangular lattice. Figure~\ref{fig:honeycomb3}(b) shows the interaction-renormalized bandstructure where the two occupied bands are polarized with the opposite spins: the dispersive band with spin $\uparrow$ is the lower band of the effective honeycomb lattice and the nearly flat band with spin $\downarrow$ is from the triangular lattice.  We show the corresponding Wannier center flow in Fig.~\ref{fig:WCF3}(b), where the constant phase is associated with the occupied spin $ \downarrow $ band on triangular lattice and the other is associated with the occupied spin $ \uparrow $ band on the honeycomb lattice winding one time across $ \bm{b}_{m_1} $.

\end{document}